\documentclass[preprint,showpacs,showkeys]{revtex4}
\usepackage{epsf,psfig,pstricks,pst-grad}
\usepackage{amsmath}
\usepackage{amssymb}
\usepackage{epic}
\usepackage[dvips]{graphicx}% Include figure files
\usepackage{dcolumn}% Align table columns on decimal point
\usepackage{bm}% bold math

\begin{document}

%\preprint{APS/123-QED}

\title{Kelly Criterion
revisited: optimal bets}% Force line breaks with \\

\author{Edward W.~Piotrowski}\email{ep@alpha.uwb.edu.pl}\homepage{http://alpha.uwb.edu.pl/ep/sj/index.shtml}
\author{ Ma\l gorzata Schroeder}%
 \email{mszuli@math.uwb.edu.pl}
 \affiliation{Institute of Mathematics,
University of Bia\l ystok, Lipowa 41, Pl 15424 Bia\l ystok,
Poland}

%\affiliation{%
%Institute of Mathematics, University of Bia\l ystok, Bia\l ystok, Poland.}%

\date{\today}% It is always \today, today,
\begin{abstract}
Kelly criterion, that maximizes the expectation value of the
logarithm of wealth for  bookmaker bets, gives an advantage over
different class of strategies. We use projective symmetries for a
 explanation of this fact. Kelly's approach
allows for an interesting financial interpretation of the
Boltzmann/Shannon entropy. A ``no-go'' hypothesis for big
investors is suggested.
\end{abstract}

%Use showkeys class option if keyword
\maketitle

\section{Introduction}
${}$\indent
When John L.~Kelly was working for Bell Labs, he observed  analogies
between calculation of the optimal player's stake
 who enters into a gambling game and the effective transmission of
 information in a noisy communication channel.
 During the last half century the strategy which was proposed by Kelly
 became very popular among gamblers
 and inspired many authors of articles and books. The original paper dated
from
 1956  is hardly available. Therefore, with the
AT\&T consent,   it has been recently reproduced in \LaTeX .
Today, strategies based on Kelly criterion  are successfully
adopted in financial markets, blackjack and even horse races. The
central problem for gamblers is to find positive expectation bets.
But the gambler also needs to know how to manage his money, i.e.
how much to bet. Application of the Kelly criterion  in blackjack
was quite successful \cite{thorp}. If all blackjack bets paid even money, had positive
expectation and were independent, the resulting Kelly betting
recipe when playing one hand at a time would be extremely simple:
bet a fraction of your current capital equal to your expectation.
Does the Kelly criterion unambiguously specify the winning
strategy? In the thermodynamic limit the maximization of the
expectation value of  logarithm of the profit rate still leaves
freedom of adopting different strategies. Because of calculational
difficulties,  only the limit case of extreme profit can be given
in a concise analytical form. Kelly's association suggests a
method  of describing
 effectiveness of agents investing in
 the financial market in thermodynamical terms.

\section{The rules of the game}
Let us consider the simplest bookmakers bet. It can be described
by disjoint alternative of two events (the random events, the
majority branches of events), which we denote $1$ and $2$. We
assume that $in_k^m$ (where $k=1,2$, and $m\in\mathbb{N}$) is the
fraction of current capital of $m$-th gambler, bet on event $k$,
and
\begin{equation}
I\negthinspace N_k:=\sum_m in^m_k>0
\label{jeden}
\end{equation}
describes the sum of wagers from all the gamblers of the bet.
Accordingly,  $out_k^m$ is the odds paid for the $m$-th gambler on
the occurrence of the $k$-th event.

The following conditions define our bet:
\renewcommand{\labelenumi}{\Alph{enumi}}
\begin{description}
\item[{\rm A)}] We shall consider the case of "fair" odds (payoff odds are calculated by sharing the pool among all placed bets
 -- the parimutuel betting), i.e.
\begin{equation}
\forall_{k,m}\,\,\, out^m_k=\, \alpha_k\, in^m_k \,,
\label{dwa}
\end{equation}
where $\alpha_k
\negthinspace\in\negthinspace \mathbb R_+$.
\item[{\rm B)}] All fees and taxes are not taken into account which means that all the money is paid out to the winners:
\begin{equation}
I\negthinspace N_1+I\negthinspace N_2=\sum_m out_1^m= \sum_m out_2^m \,
\label{trzy}
\end{equation}
\noindent (it should be noted that the gamblers are placing their bets differently, it means that the winners take all the pool).
\end{description}
The trade balance $(\ref{trzy})$ is the natural premise. Let us
observe that all costs and bookmaker's benefit might be the fee
for participation in the game. The winner  carries out an analysis
of this cost after the winnings. The above condition A is
equivalent to the statement that the bookmaker bet is a good offer
on the effective market without an opportunity of arbitrage
between the gamblers.

The conditions A and B describe uniquely the value of the factors
$\alpha_k$, which can be derived from formulas
 $(\ref{jeden})$, $(\ref{dwa})$ and $(\ref{trzy})$. We have that:
$$
\forall_k\,\,\, \alpha_k=\, \frac{I\negthinspace N_1+I\negthinspace N_2}{I\negthinspace N_k}\,.
$$

The formal description of the bookmaker bets with majority
of branches of events might be created hierarchically as the binary
tree with the leafs -- elementary events, e.g\mbox{.} by analogy
to the construction of tree-shaped key to
compressing/decompressing Huffman code \cite{cormen}. It follows
that our binary bet is {\em universal}\/, i.e\mbox{.} many kinds of financial decisions
we can describe as the systems based on a hierarchy of formal binary bets.
 Within the analogical model for
insurance, the differences would only appear in the equation
$(\ref{roznice})$ of balanced benefit. In this case the balance
$(\ref{roznice})$ includes the possible loses which are relevant
in insurances.

\section{The average gambler's gain}
We will omit the subscript $m$ because  we analyze the gain of a
particular gambler. We will use the following notation: $all_0$ --
the gambler's capital before placing bets, accordingly $all_1$
-- the gambler's capital after result of $1$, analogically $all_2$
-- after result of $2$. The balance of expense and gambler's
income is given by the formula:
\begin{equation}
all_k = all_0 - in_1 -in_2 + out_k\,.
\label{roznice}
\end{equation}
From the  projective geometry point of view, where the assets
exchange are described without  scale effect in a natural way, the
profit (up to a multiplicative constant) is the unique additive
invariant of the  group of homographies, which include all
objective irrelevant transformation between different ways of
mathematical modelling of the financial effect. When the $k$-th
event occurs the bookmaker bet, in this context, is represented by
the following configuration of the straight lines \footnote{Or the
dual configuration.}:
\begin{center}
\setlength{\unitlength}{0.0006in}
\begingroup\makeatletter\ifx\SetFigFont\undefined%
\gdef\SetFigFont#1#2#3#4#5{%
  \reset@font\fontsize{#1}{#2pt}%
  \fontfamily{#3}\fontseries{#4}\fontshape{#5}%
  \selectfont}%
\fi\endgroup%
{\renewcommand{\dashlinestretch}{30}
\begin{picture}(6987,5667)(0,-10)
\drawline(749,1158)(6975,161)
\drawline(1097,710)(649,5391)
\drawline(998,710)(1795,5421)
\drawline(649,959)(6577,2950)
\dashline{60.000}(500,3149)(6875,62)
\dashline{60.000}(450,5291)(4684,12)
\dashline{60.000}(500,5640)(2044,361)
\put(50,2870){\makebox(0,0)[lb]{{\SetFigFont{12}{14.4}{\rmdefault}{\mddefault}{\updefault}\scriptsize portfolio }}}
\put(-350,2700){\makebox(0,0)[lb]{{\SetFigFont{12}{14.4}{\rmdefault}{\mddefault}{\updefault}\scriptsize with coupons}}}
\put(1125,870){\makebox(0,0)[lb]{{\SetFigFont{12}{14.4}{\rmdefault}{\mddefault}{\updefault}$0$}}}
\put(6260,100){\makebox(0,0)[lb]{{\SetFigFont{12}{14.4}{\rmdefault}{\mddefault}{\updefault}$\infty$}}}
\put(6075,2880){\makebox(0,0)[lb]{{\SetFigFont{12}{14.4}{\rmdefault}{\mddefault}{\updefault}$u$}}}
\put(750,4910){\makebox(0,0)[lb]{{\SetFigFont{12}{14.4}{\rmdefault}{\mddefault}{\updefault}$\infty'$}}}
\put(5550,200){\makebox(0,0)[lb]{{\SetFigFont{12}{14.4}{\rmdefault}{\mddefault}{\updefault}$m$}}}
\put(600,3900){\makebox(0,0)[lb]{{\SetFigFont{12}{14.4}{\rmdefault}{\mddefault}{\updefault}$n$}}}
\put(1500,4946){\makebox(0,0)[lb]{{\SetFigFont{12}{14.4}{\rmdefault}{\mddefault}{\updefault}$w$}}}
\put(3920,350){\makebox(0,0)[lb]{{\SetFigFont{12}{14.4}{\rmdefault}{\mddefault}{\updefault}$all_0$}}}
\put(1520,720){\makebox(0,0)[lb]{{\SetFigFont{12}{14.4}{\rmdefault}{\mddefault}{\updefault}$all_k$}}}
\end{picture}
}
\end{center}
The straight lines $u$ and $w$ \footnote{On the market of goods,
the lines $w$ and $u$ represent the hyperplane of codimension
one.} express the proportion  of  an exchange (the market rate) of
the initial capital on the  our  financial obligations and
analogically the final obligation, respectively. The lines $m$ and
$n$ denote the portfolio with ready money (before the closing a
business and after the settlement of accounts of the bets) and the
portfolio which include the bookmaker coupons (when the bet has
been in effect). The set of the projection points $\{m,n\}$ is the
unique invariant of the game which is defined by the gambler's
strategy. The unique representation of the exchange of the
bookmaker stakes $u$ and $w$ is possible only with the accuracy of
the homograpic transformation. Thus  bookmaker stakes are the
covariant components of the model. They depend on choice  of the
basis of goods units (that means the basis of vector space which
is related to parametric portfolios -- the projection points of
homogeneous coordinates). Thus the set $\{m,n\}$, often  called as
an absolute, allows one to equip  the projective space with the
Hilbert metrics \cite{buseman} and non-arbitrary  measure of the
distance between two portfolios $u$ and $w$ given by this metrics.
It represents the profit flow in the transaction cycle
$m\stackrel{u}{\rightarrow} n\stackrel{w}{\rightarrow} m$\/. This
profit is equal to \cite{piotrowski,piotrsl}:
$$
z_k:=\ln |[n,u,w,m]|=\ln all_k - \ln all_0\,,
$$
where $[n,w,u,m]$ is a cross ratio of the projective points $n$, $u$, $w$, and $m$. Let us denote the percentage share of gambler's capital in both cases of the bookie bets by $l_k:=\tfrac{in_k}{all_0}$ and let $p_k$ be the probability of the $k$'th event. If $\left|in_k\right|\leq\negthinspace I\negthinspace N_k$ then the gambler's expected profit is equal to:
\begin{equation}
E(z_k)(l_1,l_2):= p_1 z_1+p_2 z_2 = p_1 \ln (1+\tfrac{I\negthinspace N_2}{I\negthinspace N_1} l_1-l_2)+p_2 \ln (1+\tfrac{I\negthinspace N_1}{I\negthinspace N_2} l_2-l_1)\,.
\label{zys}
\end{equation}
\section{Maximal  expected growth rate of wealth}
The gambler bets the stakes  $\bar{l}_1$ and $\bar{l}_2$ such that
her/his expected profit is the maximal one:
$$
E(z_k)(\bar{l}_1,\bar{l}_2):=\max_{l_1,l_2} \{E(z)(l_1,l_2)\}\,.
$$
By using the standard method we find the extremum  of the differentiable function and we obtain that the family $(\bar{l}_1,\bar{l}_2)\negthinspace\in\negthinspace\mathbb{R}^2$ of the strategies solutions of above problem is described by the following straight line equation:
\begin{equation}
(\bar{l}_1-p_1)\,I\negthinspace N_2=(\bar{l}_2-p_2)\,I\negthinspace N_1\,,
\label{prosta}
\end{equation}
and the maximal profit is given by:
\begin{equation}
E(z_k)(\bar{l}_1,\bar{l}_2)= -\sum_{k=1,2} p_k \ln \tfrac{I\negthinspace N_k}{I\negthinspace N_2+I\negthinspace N_2}-S\,,
\label{zysk}
\end{equation}
where  $S=-\sum_k p_k \ln p_k$ is Boltzmann/Shannon entropy.
Thanks to this Eq\mbox{.} $(\ref{zysk})$, we have the financial interpretation  of Kelly's
formula. The maximal profit given by Eq\mbox{.} $(\ref{zysk})$ has two
components. The first of these elements is the profit on
unpopularity of the winning bet (the seer's profit) $-\sum_{k=1,2}
p_k \ln \tfrac{I\negthinspace N_k}{I\negthinspace
N_2+I\negthinspace N_2}\,$, and second means the (minus) entropy
$-S$ of the branching. The value of $E(z)(\bar{l}_1,\bar{l}_2)$ is
nonnegative -- a rational gambler cannot loose. Thus her/his average profit equals to $0$ if the resultant
preferences adopt to the probability measurement to the
branching: $p_1 I\negthinspace N_2 = p_2 I\negthinspace N_1$.
Consequently, one can make profit in the bookie bet only when
somebody bets irrationally in the same game.

\section{The optimal strategy}
Till this moment we have assumed that there is no any additional
condition for the simplest bookmaker bet, we allow the short
position of the gamblers (negative value of $\bar{l}_k$). This is
the reason why the rational gambler has the freedom of choosing
the value of financial outlays $\bar{l}_1\negthinspace+\bar{l}_2$
which is placed in bookmaker bets. In the absence of short
positions (a typical restriction on the bet $l_1,l_2\geq 0$) we
assume that the rational gambler diversificates the risk in such a
way that she/he bets only the minimal part of their resources.
From all the strategies $(\ref{prosta})$ we choose the optimal
one:
$$(l_1^\ast\negthinspace =\negthinspace p_1\negthinspace - \negthinspace \tfrac{I\negthinspace N_1}{I\negthinspace N_2}\,p_2,\,l_2^\ast\negthinspace =\negthinspace 0)\,,
$$
when $p_1I\negthinspace N_2\negthinspace >\negthinspace
p_2I\negthinspace N_1$, or, equivalently, the one that can be obtained
by the transposition  $1\negthinspace\leftrightarrow\negthinspace
2$ of the indices $k$.

If we do not have the information about proportion  $\tfrac{I\negthinspace N_1}{I\negthinspace N_2}$ then we use Laplace's Principle of Indifference ($I\negthinspace N_1\negthinspace=\negthinspace I\negthinspace N_2$), and in this case (when $p_1\negthinspace>\negthinspace p_2$) the optimal stakes are ($(l_1^\ast\negthinspace =\negthinspace p_1\negthinspace - \negthinspace p_2,\,l_2^\ast\negthinspace =\negthinspace 0)$, see~[Kelly].

\section{Big gamblers -- ``no-go'' hypothesis}

Let us now consider the variant of the binary bet when the gambler's contribution of the $in_k$ to the sum $IN_k$ is not neglected. If the gambler pays to the pool, the pool of the bets grows from $I\negthinspace N_1 + I\negthinspace N_2$ to
$(1+\delta)(I\negthinspace N_1 + I\negthinspace N_2)$, where $\delta\negthinspace\in\negthinspace\mathbb{R}$. Consequently the parts of the pool corresponding to different events are going to change from  $I\negthinspace N_k$ to
$I\negthinspace N_k +\, \delta \tfrac{l_k}{l_1+l_2}(I\negthinspace N_1 +I\negthinspace N_2)$.Then the part of the gambler's expected profit $E_\delta(z)(l_1,l_2)$ which is linear in $\delta$ will be given by:
$$
E_\delta(z)(l_1,l_2) = p_1 \ln (1+\tfrac{\tfrac{I\negthinspace N_2}{I\negthinspace N_1+I\negthinspace N_2}+\delta\tfrac{l_2}{l_1+l_2}}{\tfrac{I\negthinspace N_1}{I\negthinspace N_1+I\negthinspace N_2} +\delta\tfrac{l_1}{l_1+l_2}}\, l_1-l_2)+p_2 \ln (1+\tfrac{\tfrac{I\negthinspace N_1}{I\negthinspace N_1+I\negthinspace N_2}+\delta\tfrac{l_1}{l_1+l_2}}{\tfrac{I\negthinspace N_2}{I\negthinspace N_1+I\negthinspace N_2}+\delta\tfrac{l_2}{l_1+l_2}}\, l_2-l_1)=
$$
$$E(z)(l_1,l_2)+\frac{\partial E_\delta(z)(l_1,l_2)}{\partial \delta}\biggr\vert_{\delta=0}\hspace{-.6em}\delta+O[\delta]^2\,,
$$
where
\begin{equation}
\frac{\partial E_\delta(z)(l_1,l_2)}{\partial \delta}\biggr\vert_{\delta=0}=
\tfrac{\tfrac{l_1}{l_1+l_2}\,\tfrac{I\negthinspace N_1+I\negthinspace N_2}{I\negthinspace N_1}}{\tfrac{I\negthinspace N_1}{l_2 I\negthinspace N_1-l_1 I\negthinspace N_2}\,-1}\,p_1 +
\tfrac{\tfrac{l_2}{l_1+l_2}\,\tfrac{I\negthinspace N_1+I\negthinspace N_2}{I\negthinspace N_2}}{\tfrac{I\negthinspace N_2}{l_1 I\negthinspace N_2-l_2 I\negthinspace N_1}\,-1}\,p_2\,.
\label{for1}
\end{equation}
It is sufficient to restrict oneself to the case when $\delta$ is
an infinitely small number and then we can consider the corrected
parameters $I\negthinspace N_1$ and $I\negthinspace N_2$ (change
of $\delta$). The extremal strategy is defined by the set of
equations:
\begin{equation}
\frac{\partial \bigl( E(z)(l_1,l_2)+\frac{\partial E_\delta(z)(l_1,l_2)}{\partial \delta}\bigr\vert_{\delta=0}\delta\bigr)}{\partial l_k}=0\,.
\label{for2}
\end{equation}
The solutions of these equations are the roots of two polynomials
of degree five \footnote{We do not give  their explicit form,
because they can be easily generated by using the set of
equations $(\ref{zys})$, $(\ref{for1})$, and $(\ref{for2})$, and
taking advantage of ~the language symbolical calculation {\em
Mathematica}\/. Few exemplary lines are added in Appendix}.
According to the fundamental theorem of Galois theory, we can
conclude that an analytic form of the conditions for the optimal
big player's strategy is not countable. We can find the family of
parameters $(\bar{l}_1,\bar{l}_2)$  by using  numerical methods,
but we never can investigate the behavior of characteristics of
the rational big gamblers.

Due to the universality of the binary bet model, we can conclude
that any type of analysis of big investors strategies, whose
appearance will disturb the financial market, will not be satisfactory 
because of principal mathematical reasons. It is also possible that macroeconomic
thermodynamics considered as the analysis of the market disturbing
strategies is forbidden by mathematics! In these contexts the
tendency of diversification in the investment might be perceived
as an escape of the investors from the unsolvable problems.

Is it really true that {\em Small Is Beautiful}\/ \footnote{This is a title of the cult book of Fritz Schumacher.} also in the markets?
\begin{acknowledgements} 
We are greatly indebted to prof.~Zbigniew Hasiewicz for helpful remarks.
This paper has been supported
by the Polish Ministry of Scientific Research and Information Technology
under the (solicited) grant No PBZ-MIN-008/P03/2003.
\end{acknowledgements} 

\section*{Appendix}
The exemple code given below is written in {\em Mathematica
5.2}\/ language and it allow one to calculate the algebraic
expressions as a nonfactorizable polynomials of degree $5$ in
variables $l_1$ i $l_2$. Some zeros of these polynomials
characterize  optimal strategies of betting the stakes in our
model of the bookmaker bet.
\begin{figure}[h]
\begin{center}
\includegraphics[height=15cm, width=15cm]{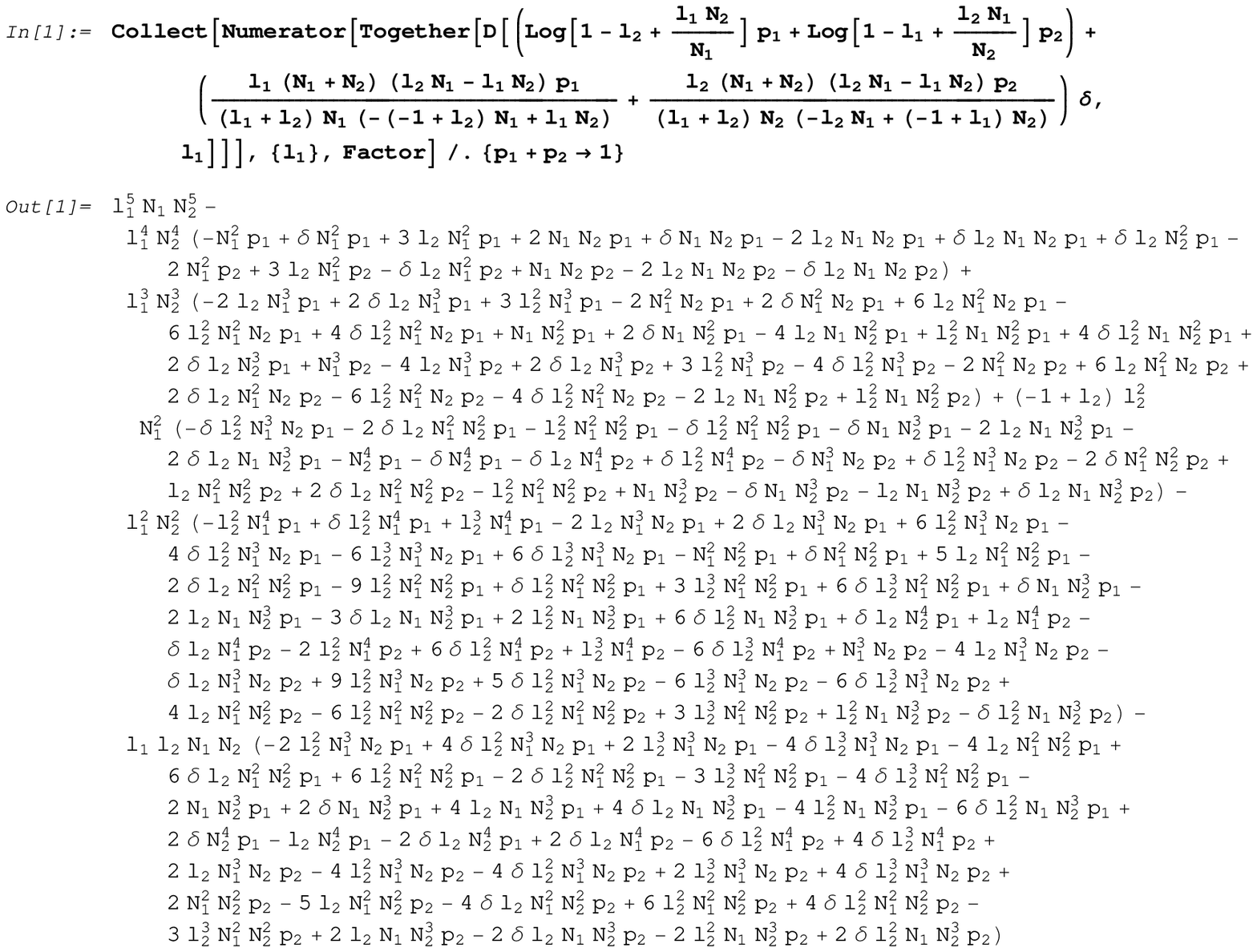}
\end{center}
\end{figure}
The second polynomial can be obtained by the transposition
$1\negthinspace\leftrightarrow\negthinspace 2$ of the indices $k$.

\end{document}